\documentclass[global]{svjour}
\usepackage{graphics}
\usepackage{amsmath}
\usepackage{subfigure}
\begin{document}
\title{Dynamics of blueshifted floating pulses in gas filled hollow-core photonic crystal fibers}
\author{M. Fac\~ao\inst{1} \and M. I. Carvalho\inst{2}}

\institute{Departamento de F\'{i}sica, Universidade de Aveiro, I3N,
3810-973 Aveiro, Portugal \and DEEC/FEUP and INESC TEC, Universidade
do Porto, Rua Dr. Roberto Frias, 4200-465 Porto, Portugal}

%
%
\titlerunning{Dynamics of blueshifted floating pulses}
\maketitle
\begin{abstract}
Frequency blueshifting was recently observed in light pulses
propagating on gas filled hollow-core photonic crystal fibers where
a plasma has been produced due to photoionization of the gas. One of
the propagation models that is adequate to describe the actual
experimental observations is here investigated. It is a nonlinear
Schr\"odinger equation with an extra term, to which we applied a
self-similar change of variables and found its accelerating
solitons. As in other NLS related models possessing accelerating
solitons, there exist asymmetrical pulses that decay as they
propagate in some parameter region that was here well defined.

\end{abstract}
\section{Introduction}
Hollow-core photonic crystal fibers (HC-PCF) have recently attracted
a lot of attention particularly because they offer ultra-long single
mode interaction lengths for nonlinear optics in gaseous media \cite{travers11}.
Among the kind of HC-PCFs, we refer the kagom\'e type that was first
reported in 2002 \cite{benabid02} and owes its name to the kagom\'e
lattice-cladding. The kagom\'e HC-PCF provides several hundred
nanometers band of guidance at low loss levels ($\approx$ 1~dB/m) and
exhibits weak anomalous group-velocity dispersion (GVD) over the
entire transmission window, with low dispersion slope. Whenever
filled with gas, it enables the self-compression of pulses to an
extent that the peak intensities may attain values above the ionization threshold
of the gas, allowing the production of a plasma. Then, the
interaction of laser light with the plasma leads to new nonlinear
effects such as the blueshifting \cite{hoelzer11} of the central wavelength of the pulse.

Recently, Saleh \textit{et al} \cite{saleh11,saleh11pra} presented
an amenable model for describing the interaction between the optical
pulse and plasma on those gas-filled kagom\'e HC-PCFs. The model was
used by them to predict the extent of frequency blueshifting by
means of a perturbation approach. Following these preliminary
perturbation results, a thorough study on the existence of
accelerating solitons to such a model, including the plasma but also
the stimulated Raman scattering term, was performed \cite{facao13}.
There, it was shown that indeed there are self-similar pulse
solutions of such model that accelerate while its central frequency
blueshifts. However, it was reported that for large strength of the
plasma nonlinearity or small pulse amplitudes, the pulses have
distinguished long tails and decay as they propagate. The latter
analysis was done with a scaling that resulted in a ODE with four parameters, one for the intensity threshold, other for the acceleration and both
the plasma and Raman terms accomodated with only one parameter and a ratio
between the strength of them . This scaling was such that
both the extra nonlinear parameter and pulse amplitude of the final
normalized model should be very small in order to describe the
actual physical experiments. Here, we disregard the Raman term and use an approximate
equation of this model that was already used in \cite{saleh11}, and obtain a
two parameter ordinary differential equation (ODE) for the
accelerating pulse profiles, finding the parameters ranges for
symmetrical pulse profiles and for long tail pulses. Furthermore,
our results, confirmed by numerical simulations of the evolution of
sech pulses, also show that even though there is always pulse decay
for the parameter range corresponding to asymmetrical pulse
profiles, the decay rate can increase or decrease during propagation
depending on these parameters.

\section{One parameter ODE and perturbation approach}
The model introduced in \cite{saleh11,saleh11pra} for pulse propagation in
nonlinear gaseous media presenting Kerr and plasma nonlinearities
has the following dimensionless version
\begin{equation}
i\frac{\partial q}{\partial\xi}+\frac{1}{2}\frac{\partial^2 q}{\partial \tau^2}
+|q|^2q-\phi_Tq\left(1-\text{e}^{-\sigma\int_{-\infty}^\tau\Delta|q|^2\Theta(\Delta|q|^2)d\tau'}\right)=0
\label{pde}
\end{equation}
where $q=(\gamma z_0)^{1/2}\psi$ represents the optical field
envelope $\psi$, $\xi=Z/z_0$ and $\tau=t/t_0$ are normalized
versions of the propagation distance $Z$ and time $t$ in a reference
frame that travels at group velocity,
$\phi_T=\tfrac{1}{2}k_0z_0(\omega_T/\omega_0)^2$ refers to the
maximum plasma frequency $\omega_T$ and
$\sigma=\tilde{\sigma}t_0/(A_\text{eff}\gamma z_0)$ to the
photoionization cross-section $\tilde{\sigma}$,
$\Delta|q|^2=|q|^2-|q|^2_\text{th}$ where $|q|^2_\text{th}$ is
related with the ionization intensity threshold and $\Theta$ is the
Heaviside step function. On these latter relations,
$z_0=t_0^2/|\beta_2|$ is the so-called dispersion length ($\beta_2$
is the GVD parameter), $t_0$ is an arbitrary time chosen similar to
the pulse duration, $A_\text{eff}$ is the effective optical mode
area, $\gamma$ is the nonlinear Kerr parameter, $\omega_0$ is the
central frequency of the pulse and $k_0$ the corresponding vacuum
wavenumber. Note that this model assumes that the recombination time
is longer than the pulse and does not consider the ionization
induced loss that is small especially for pulses whose peak is
barely above the threshold.

Referring to experimental data \cite{hoelzer11}, we arrived to
values of $\sigma$ around 10$^{-4}$ and values for $q$ and
$q_\text{th}$ around the unity. Hence, the exponential term may be
approximated by a two term Maclaurin expansion giving
\begin{equation}
i\frac{\partial q}{\partial\xi}+\frac{1}{2}\frac{\partial^2 q}{\partial \tau^2}
+|q|^2q-\eta q\int_{-\infty}^\tau\Delta|q|^2\Theta(\Delta|q|^2)d\tau'=0
\label{pde-2}
\end{equation}
where $\eta=\sigma\phi_T$. Applying the same accelerating variable
as in \cite{facao13}, namely, $T=\tau+\frac{a}{4}\xi^2+b\xi$ and
$q(\xi,\tau)=F(T)\exp(i\theta(\xi,T))$, with $F$ and $\theta$ real,
we obtain the ODE for $F$
$$F''+\left(aT-D+2F^2-2\eta\int_{-\infty}^T\Delta F^2\Theta(\Delta F^2)dT'\right)F=0$$
and the phase
\begin{equation}\theta(\xi,T)=-\left(\frac{a}{2}\xi+b\right)T+\frac{1}{2}(D+b^2)\xi+\frac{1}{4}ba\xi^2+\frac{1}{24}a^2\xi^3+E.
\label{phase}
\end{equation}
In this case, the ODE for the pulse profile may be further simplified if one uses the following change of variables
$$F(T)=\frac{2\eta}{\beta}P(\zeta),\quad \zeta=\frac{2\eta}{\beta}T,$$
obtaining
\begin{equation}
P''+\left(\beta\zeta-G+2P^2-\beta\int_{-\infty}^\zeta\Delta P^2\Theta(\Delta P^2)d\zeta'\right)P=0,
\label{ode-P}
\end{equation}
where $a$ and $\eta$ were replaced by the single parameter
$\beta=\left(8\eta^3/a\right)^{1/2}$
and $G=(\beta/2\eta)^2D$ is the new constant.

As the evolution equation (\ref{pde-2}) is the nonlinear
Schr\"odinger equation (NLSE) with the extra term representing the
plasma-light interaction, sech soliton solutions may be anticipated for the ODE (\ref{ode-P}) with $\beta=0$. In fact,
these sech solutions are of the form
$P_0(\zeta)=\sqrt{G}\text{sech}\left[\sqrt{G}(\zeta-\zeta_0)\right]$.
Hence, let us write an expansion for $P(\zeta)$ in the form
$$P(\zeta)=P_0(\zeta-\zeta_0)+\beta P_1(\zeta)+\cdots$$
and introduce it into (\ref{ode-P}) which yields
$$P_1''+(-G+6P_0^2)P_1=-\zeta P_0+P_0\int_{-\infty}^\zeta (P_0^2-P_\text{th}^2)\Theta(P_0^2-P_\text{th}^2)d\zeta'$$
The homogeneous part of the above equation is satisfied by $P_0'$ so that the solvability condition is
$$\int_{-\infty}^\infty\left(-\zeta P_0P_0'+P_0P_0'\int_{-\infty}^{\zeta} (P_0^2-P_\text{th}^2)\Theta(P_0^2-P_\text{th}^2)d\zeta'\right)
d\zeta=0$$ which gives the following algebraic equation for the
amplitude squared, $G$, of the zero-order soliton solution
\begin{equation}
G=\frac{3}{2}\left(1-\frac{P_\text{th}^2}{G}\right)^{-3/2}.
\label{eq-G}
\end{equation}
As long as $\beta$ is relatively small, we may say that the peak
amplitude of $q$ is $|q|_\text{peak}=\frac{2\eta}{\beta}\sqrt{G}$,
which implies that the acceleration parameter is
$a=\frac{2\eta}{G}|q|_\text{peak}^2$ and the frequency shift becomes
$\Delta\omega=-\frac{d\theta}{dT}=\frac{\eta}{G}|q|_\text{peak}^2\xi$.
Note that in case the intensity threshold is zero, the equation
(\ref{eq-G}) may be exactly solved to $G=3/2$, which gives
$a=\frac{4\eta}{3}|q|_\text{peak}^2$ and
$\Delta\omega=\frac{2\eta}{3}|q|_\text{peak}^2\xi$, indicating that
the blueshift is proportional to the square of the peak amplitude
\cite{saleh11}. On the other hand, for finite $|q|_\text{th}$, the
use of (\ref{eq-G}) allows us to write
$a=\frac{4\eta}{3|q|_\text{peak}}\left(|q|_\text{peak}^2-|q|_\text{th}^2\right)^{3/2}$
and $\Delta\omega=\frac{2\eta}{3q|_\text{peak}}
\left(|q|_\text{peak}^2-|q|_\text{th}^2\right)^{3/2}\xi$,
expressions that clearly show that the actual dependence of the
blueshift process on the peak amplitude is more complex, with both
the pulse peak amplitude and the threshold intensity playing an
important role in it \cite{facao13}.

\section{Pulse profiles}
The pulse solutions to ODE (\ref{ode-P}) were obtained using a
shooting method that relied on Airy function asymptotics and first
estimates coming from the above perturbation approach. In fact, at
the tails the eq. (\ref{ode-P}) reduces to
$$P''+(\beta\zeta-G-\beta\Lambda_\infty)P=0,$$
where $\Lambda_\infty=0$ if $\zeta\rightarrow -\infty$ and
$\Lambda_\infty=\int_{-\infty}^\infty \Delta {P^2}\Theta (\Delta
{P^2})d\zeta'$ if $\zeta\rightarrow \infty$, which is equivalent to
an Airy equation by the following change of variables
$z=-\beta^{1/3}(\zeta-\Lambda_\infty)+\beta^{-2/3}G$. Hence, the
asymptotics of the pulses conform with Airy functions,
particularly, they match the Ai$(z)$ at the left tail and
Bi$(z)$ at the right tail since these are the exponentially decaying
Airy solutions at $z\rightarrow\infty$ and $z\rightarrow 0^+$,
respectively. Note that this is true as long as $z$ is in the
positive semi-axis.

\begin{figure}
\resizebox{0.75\textwidth}{!}{%
\includegraphics{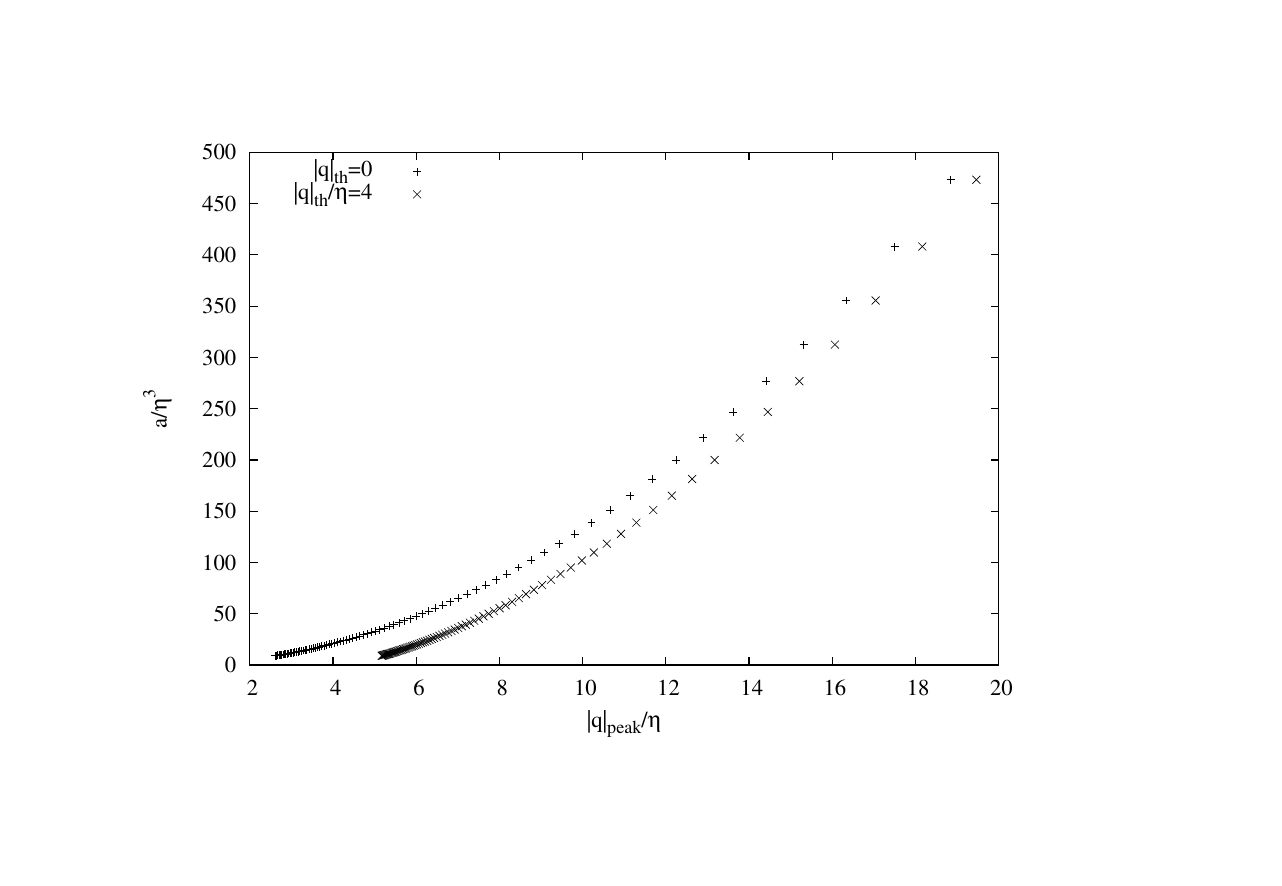}}
\caption{\label{acceleration}Dependence of acceleration
parameter on the pulse peak amplitude, normalized by $\eta^3$ and $\eta$, respectively, for $|q|_\text{th}=0$ and $|q|_\text{th}/\eta=4$.}
\end{figure}

\begin{figure}
\resizebox{0.75\textwidth}{!}{%
\includegraphics{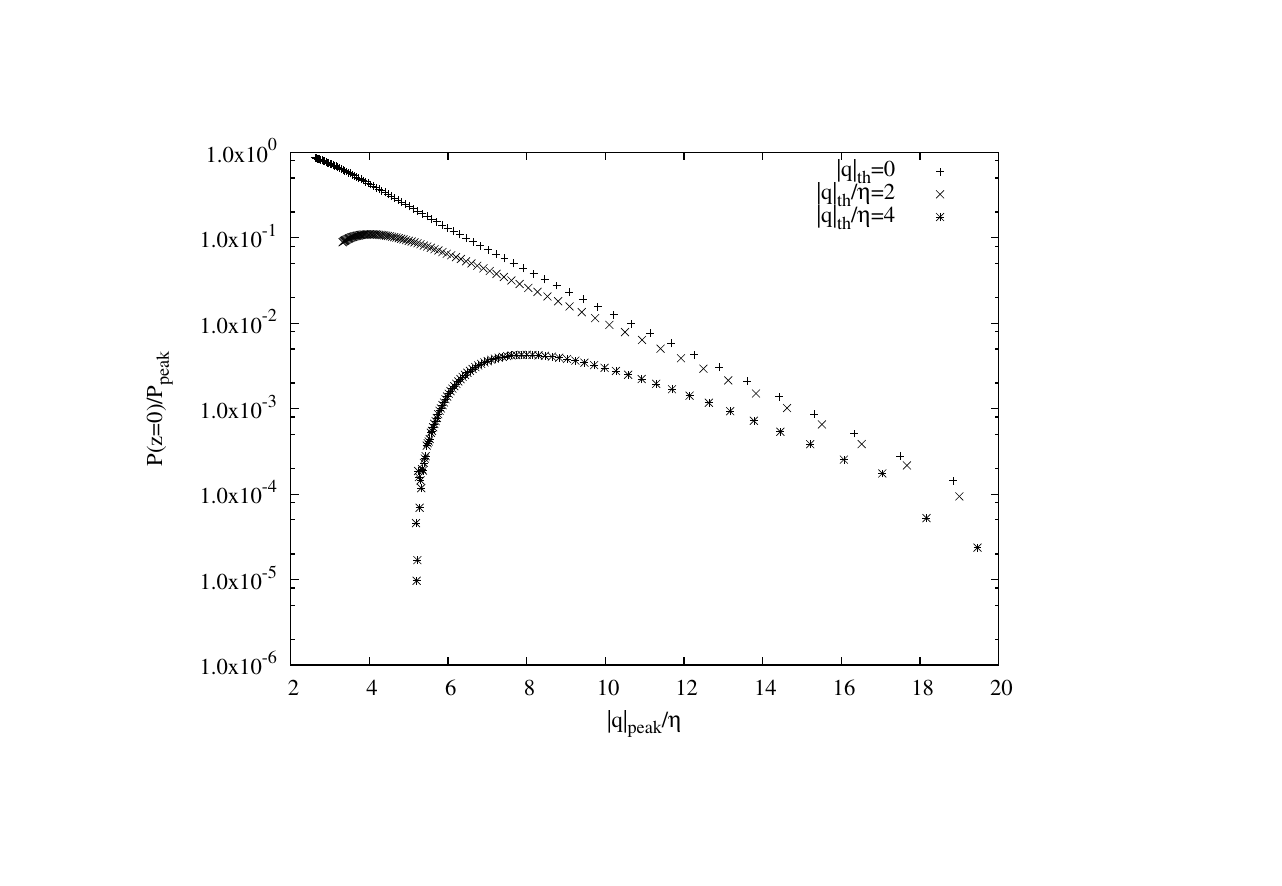}}
\caption{\label{localization}Ratio $P(z=0)/P_\text{peak}$ against $|q|_\text{peak}/\eta$ for several $|q|_\text{th}$.}
\end{figure}

The analysis up to this stage may seem very similar to the analysis
applied to the more general equation (\ref{pde}) in \cite{facao13}.
Also, the results obtained there could possibly predict the profile
characteristics of pulse solutions of (\ref{pde-2}), however, the
shooting calculations with the ODE obtained in \cite{facao13} and
small values of peak amplitudes and plasma strengths would involve
Airy function values for large arguments that are not easily
evaluated. Moreover, the present ODE is, for each $P_\text{th}$, a
one parameter ODE, thus, we may obtain all the amplitudes and
accelerations with one run that spans a considerable range of
$\beta$. The acceleration results for two of these runs are in
Fig.~\ref{acceleration}. Regarding the pulse profile
characteristics, our results confirm that the asymmetry will happen
for small peak amplitudes or large plasma term strength. The long
right tails that characterize the asymmetry of these pulses are
related with the $\zeta$ location of the pulse, namely, long tails
are expected if $\zeta_0$ is such that the corresponding $z$ of the
Airy equation is close to zero. In such cases, the asymptotics at
the right tail cannot be exponentially since, at $z<0$, Bi$(z)$ is
no longer exponentially decaying but instead both Bi and Ai are
algebraically decaying functions. In order to visualize the ranges
of peak amplitude, threshold amplitude and $\eta$ values for which
the long tails exist, we have plotted the ratio of the value of $P$
at $\zeta(z=0)$ to $P_\text{peak}$ against $|q|_\text{peak}/\eta$ for
several $|q|_\text{th}$ (Fig.~\ref{localization}). The larger this
ratio is, the closer to $z=0$ is the pulse location, and thus the
longer are the right tails and the asymmetry. The analysis of
results presented in Fig.~\ref{localization} shows that for
$q_\text{th}=0$, the long tails happen for small
$|q|_\text{peak}/\eta$, achievable with small peak amplitudes and
relatively large plasma strength (recall that we are dealing with an
approximation of the initial model that is valid for small plasma
term). In the more realistic case of $q_\text{th}\neq 0$, the long
tails happen in the same conditions with the exception that the
profile is closest to the Airy $z=0$ for some small finite
$|q|_\text{peak}/\eta$.

\begin{figure}
\resizebox{0.75\textwidth}{!}{%
\includegraphics{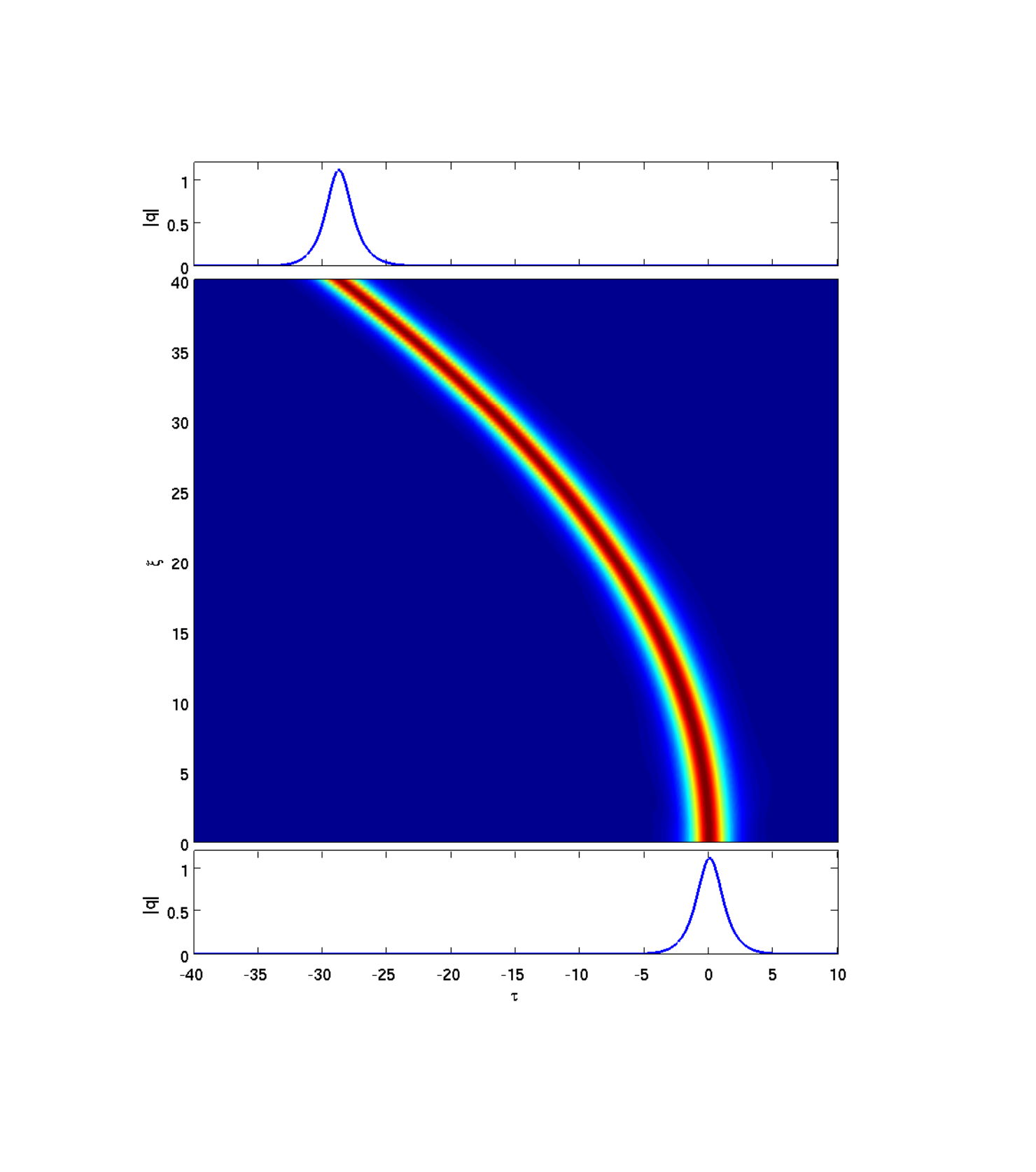}}
\caption{\label{evolution1} Pulse evolution for $|q|_\text{peak}=1.12$, $\eta=0.18$ and $q_\text{th}=0.88$.}
\end{figure}

\begin{figure}
\resizebox{0.75\textwidth}{!}{%
\includegraphics{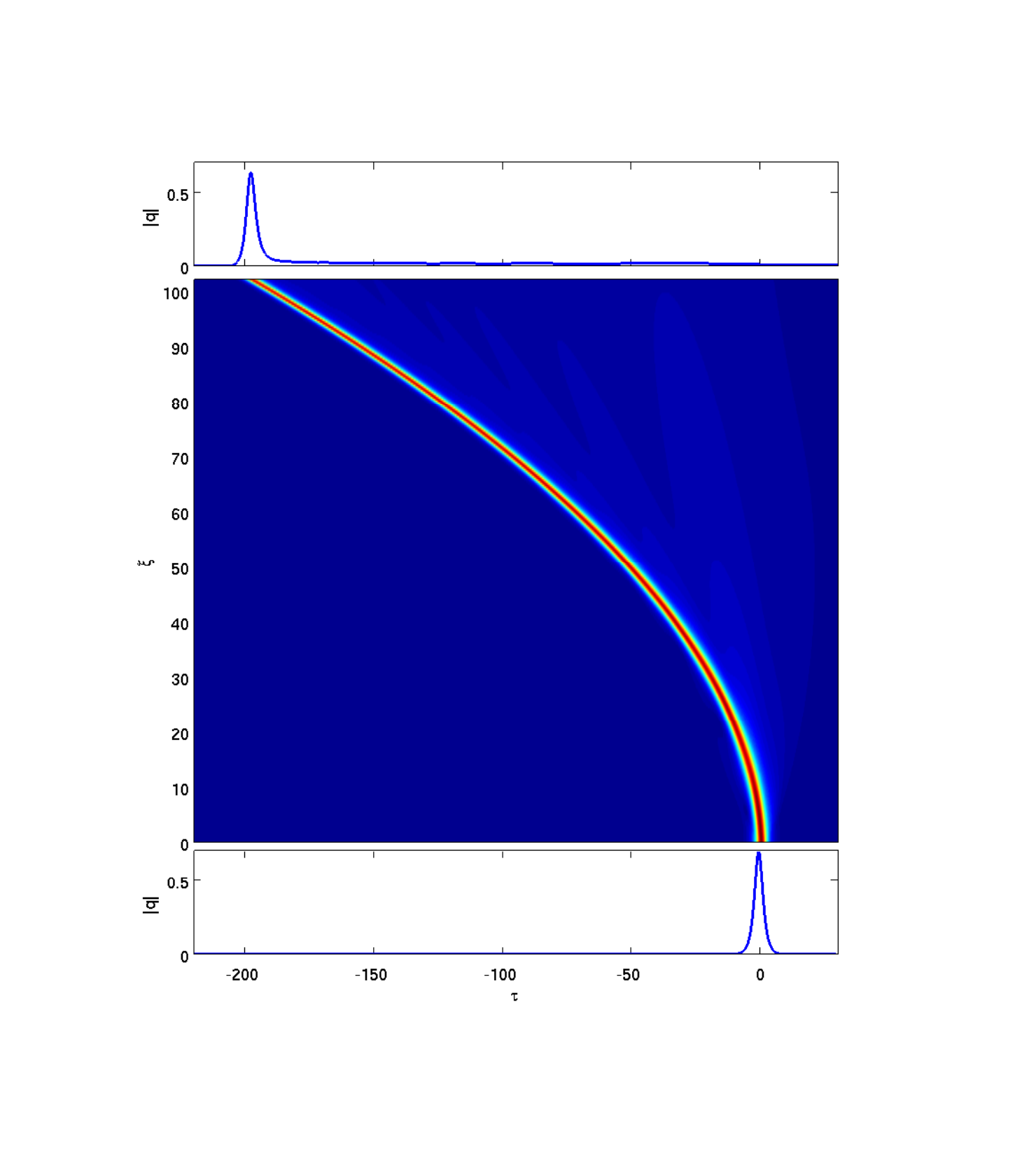}}
\caption{\label{evolution2} Pulse evolution for
$|q|_\text{peak}=0.69$, $\eta=0.35$ and $q_\text{th}=0.45$.}
\end{figure}

\begin{figure}
\resizebox{0.75\textwidth}{!}{%
\includegraphics{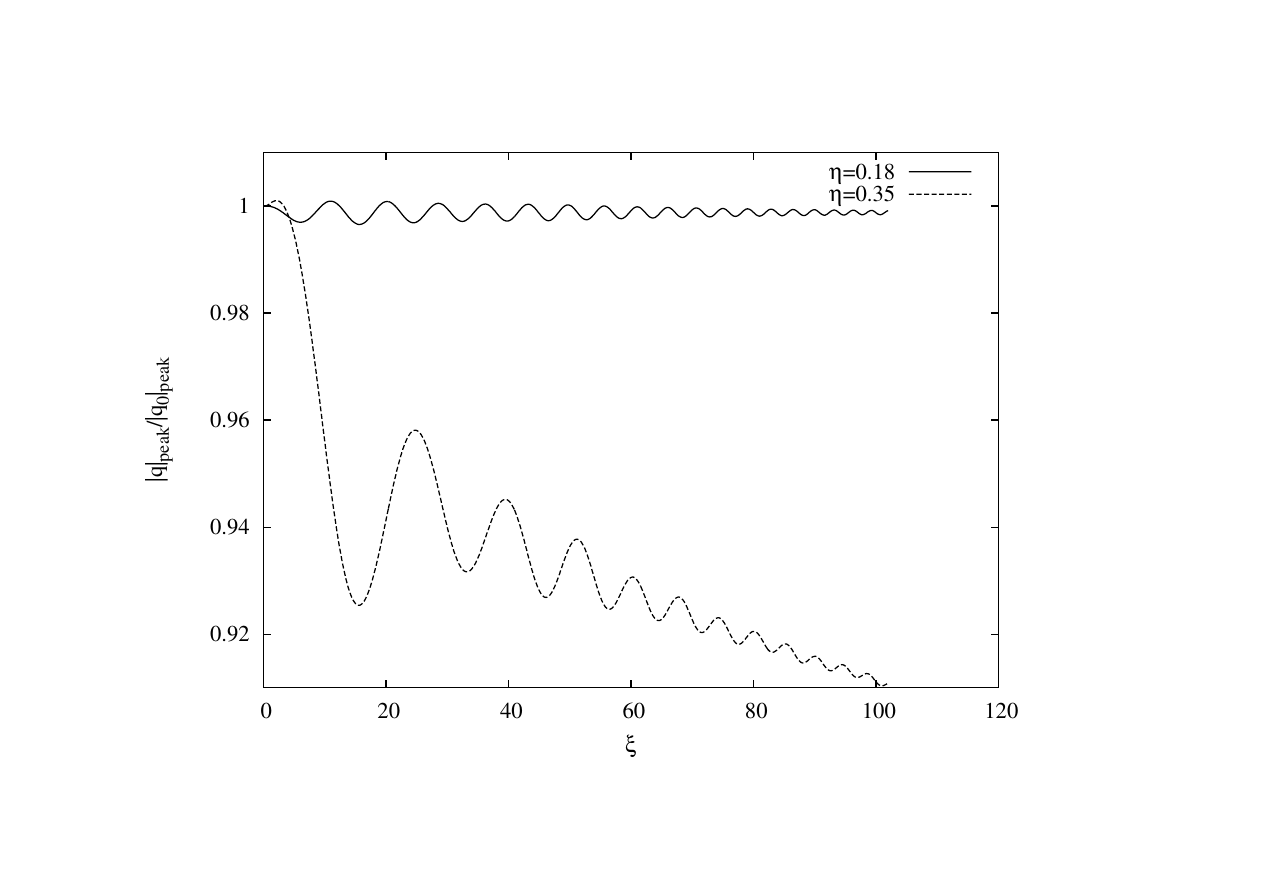}}
\caption{\label{peaks} Peak amplitude evolution for
$|q|_\text{peak}=1.12$, $\eta=0.18$ and $q_\text{th}=0.88$ ($\beta=0.78$) and for $|q|_\text{peak}=0.69$, $\eta=0.35$ and
$q_\text{th}=0.45$ ($\beta=1.9$).}
\end{figure}

In order to assess how the existence of such long tails would be
perceived in a real experiment, we decided to numerically
investigate the propagation of a fundamental soliton with peak
amplitude and duration similar to the ones that have allowed the
experimental observation of the blueshift effect \cite{hoelzer11}.
For this purpose,  equation (\ref{pde-2}) was solved using the
values presented in  Saleh \textit{et al} \cite{saleh11}
($I_\text{th}\simeq64$~TW/cm$^2$ and
$\tilde{\sigma}=1.03\times10^{-3}$~cm$^2$Hz/W) and other parameters
from H\"olzer \textit{et al} \cite{hoelzer11}. For a pulse with
$\lambda=800$~nm at the maximum compression width of 2.5~fs
propagating in 1.7~bar Argon filled kagom\'e PCF whose
$\beta_2=-7.5\times10^{-28}$~s$^2$/m and $\gamma=2.7\times
10^{-7}$~W$^{-1}$m$^{-1}$ ($n_2=1.8\times 10^{-23}$~m$^2$/W from
\cite{wang13}), we obtain $\sigma=2.15\times 10^{-4}$,
$\phi_T=8.26\times 10^2$ which gives $\eta=0.18$, $q_\text{th}=0.88$
and $|q|_\text{th}/\eta=4.9$. Hence, in this case the maximum value
of $P(z=0)/P_\text{peak}$ is very small (smaller than $0.01$) and
the respective pulse profiles should be symmetric and propagate
without radiation shedding, which is confirmed by the direct
simulation of (\ref{pde-2}) for $|q|_\text{peak}=1.12$ and $\xi=40$
(corresponding to 34~cm) that is shown in figure \ref{evolution1}.
On the other hand, we envisage an experiment in the same fiber and
same pulse width but for a pump wavelength of 1064~nm that would
give $\eta = 0.35$, $q_\text{th}=0.45$ and $q_\text{th}/\eta =
1.27$. In this case, figure \ref{localization} suggests that pulses
with small peak amplitudes should exhibit a long right tail and
decay as they propagate, as may be confirmed by the direct
simulation of a fundamental soliton with peak amplitude
$|q|_\text{peak}=0.69$ for $\xi=102$ (also corresponding to 34~cm)
shown in figure \ref{evolution2}. In effect, this pulse is also
propagating along a parabolic trajectory, but during propagation
several humps from the right tail become visible. The pulse decay is
more clearly seen in figure \ref{peaks}, which compares the
evolution of the pulses peak amplitude, normalized to their input
value, for the two cases considered. In effect, while the curve for
the first case exhibits small amplitude oscillations around a
constant value, in the second case the oscillations have a
considerable amplitude in the beginning, decreasing afterwards, but,
more importantly, the peak mean value is decreasing during
propagation. This behavior can be explained by the fact that the
stationary profile of the pulse corresponding to the parameters
considered in figure~\ref{evolution2} has a significant left tail
that carries infinite energy due to its Airy algebraic decay. As the
input pulse evolves towards this stationary profile, its peak will
naturally decrease in order to feed the tail that is forming. The
difference in the two cases considered in figures \ref{evolution1}
and \ref{evolution2} was just a change in the pump wavelength that
decreased the ratio $|q|_\text{th}/\eta$ and increased $\eta$, both
contributing for the observation of long tails and pulse decay.
However, other physical parameters, such as the gas pressure or the
fiber geometry, may change the accelerating pulse profile
characteristics imposing or not pulse decay during propagation. The
amplitude oscillations, observed in both cases, do not have a
constant period but their average period does increase with
$\beta^{8/3}/\eta^2$ as may be obtained by a simple analysis of the
spectral stability problem \cite{facao03}, namely, the oscillation
period should be close to the value of the edge of the continuous
spectrum and that edge is, in this problem, varying with
$\beta^{8/3}/\eta^2$ .

\begin{figure}
\resizebox{0.75\textwidth}{!}{%
\includegraphics{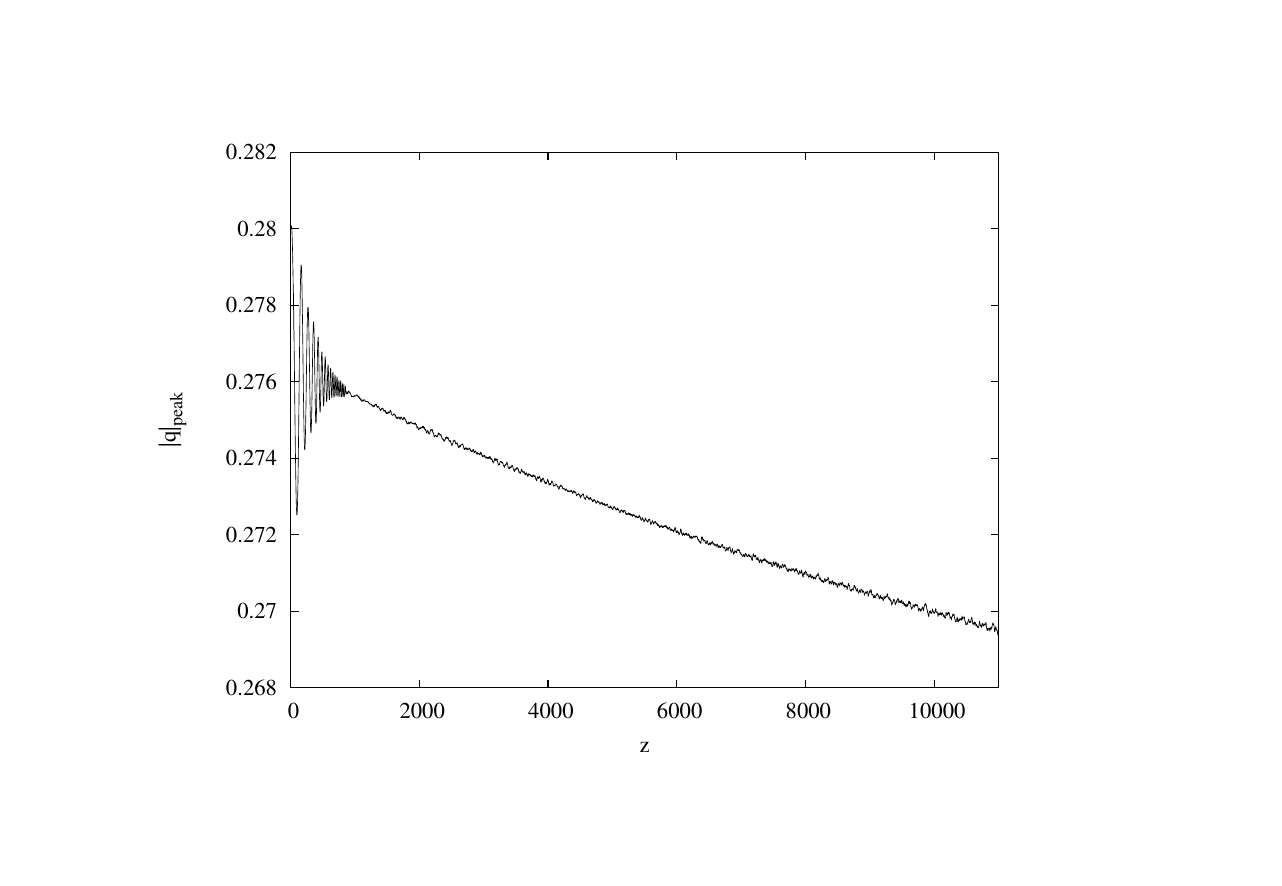}}
\caption{\label{decayrate} Peak amplitude evolution for
$|q|_\text{peak}=0.28$, $\eta=0.1$ and $q_\text{th}=0.2$ ($\beta=1.5$).}
\end{figure}

Returning to the dynamics of the decaying pulses, let us note that
since the peak amplitude decreases, the corresponding stationary
profile is also varying. In the case $q_\text{th}=0$, a smaller peak
amplitude will correspond to a pulse with a longer tail, so that,
the decay should be positively feedbacked, however, in some cases of
$q_\text{th}\neq 0$, the smaller peak amplitude may correspond to
smaller tails and we expect that the rate decay decreases. We have
confirmed this hypothesis as is shown in the simulation results of figure \ref{decayrate}, where the slope of the curve for peak amplitude decreases with distance. Moreover, taking into consideration that the plasma term will be
zero when the optical intensity drops bellow the threshold
intensity, this decay process cannot continue indefinitely.
Nevertheless, let us refer that the latter decrease in decay rate was observed even in the presence of additional loss that should have
occurred in our simulations due to numerical window limitations.

\section{Conclusions}
Starting from an evolution equation for pulse propagation in gas
filled PCFs that takes in account group velocity dispersion, Kerr
effect and relatively small plasma-light interaction, we used an
accelerating variable to find the ODE to which the pulse profiles
obey. This ODE have two parameters, the optical intensity threshold
for photoionization and one for the plasma nonlinearity strength and
acceleration which reduces the effort to compute the profiles and
accelerations. We defined two regimes of pulse profile
characteristics and propagation. There are symmetrical pulse
profiles that are very close to sech and propagate steadily, and
asymmetrical pulses presenting a long right tail that shed
radiation away as they propagate. The two regimes were identified by
two ratios, the peak amplitude over the plasma strength $\eta$ and
the photoionization threshold optical amplitude over $\eta$.

\begin{acknowledgement}

We acknowledge FCT for support through the Projects
PTDC/EEA-TEL/105254/2008 (OSP-HNLF), PTDC/FIS/112624/2009 (CONLUZ)
and PEst-C/CTM/LA0025/2011.

\end{acknowledgement}

%
%

\end{document}